\documentclass[article,10pt, twocolumn]{IEEEtran} 

\usepackage{epsf,cite,amsmath,amscd,amssymb,graphicx,latexsym,multicol}
\begin{document}
\bibliographystyle{ieeetr}

\title{\Large A Non-Cooperative Power Control Game for Multi-Carrier CDMA Systems\vspace{0.2cm}}
\author{ {Farhad Meshkati, Mung Chiang, Stuart C. Schwartz, and H. Vincent Poor}\thanks{This research was supported by the National Science Foundation under Grants
ANI-03-38807, CNS-04-17603 and CCR-04-40443.}\\
Department of Electrical Engineering, Princeton University\\Princeton, NJ 08544 USA\vspace{0cm} \\  {\textsf{\{meshkati, chiangm, stuart, poor\}@princeton.edu}}\vspace{0.4cm}  \\
Narayan B. Mandayam\\
WINLAB, Rutgers University\\Piscataway, NJ 08854 USA\vspace{0cm}\\
 {\textsf{narayan@winlab.rutgers.edu}}}

\pagestyle{empty}

\centerfigcaptionstrue

\maketitle

\thispagestyle{empty}

\begin{abstract}
In this work, a non-cooperative power control game for
multi-carrier CDMA systems is proposed. In the proposed game, each
user needs to decide how much power to transmit over each carrier
to maximize its overall utility. The utility function considered
here measures the number of reliable bits transmitted per joule of
energy consumed. It is shown that the user's utility is maximized
when the user transmits only on the carrier with the best
``effective channel". The existence and uniqueness of Nash
equilibrium for the proposed game are investigated and the
properties of equilibrium are studied. Also, an iterative and
distributed algorithm for reaching the equilibrium (if it exists)
is presented. It is shown that the proposed approach results in a
significant improvement in the total utility achieved at
equilibrium compared to the case in which each user maximizes its
utility over each carrier independently.
\end{abstract}

\section{Introduction}
Power control is used for resource allocation and interference
management in both the uplink and downlink of code division
multiple access (CDMA) systems. In the uplink, the purpose of
power control is for each user to transmit enough power so that it
can achieve the required quality of service (QoS) at the uplink
receiver without causing unnecessary interference to other users
in the system. Game theory has been used as an effective tool to
study power control for data networks (see for example
\cite{GoodmanMandayam00, Xiao01, Saraydar02, Meshkati03,
Meshkati03b, Meshkati_TCOMM}). Taking this approach, it is shown
in \cite{GoodmanMandayam00} and \cite{Saraydar02} that Nash
equilibrium is achieved when all the users aim for a target SIR,
$\gamma^*$. This is true even when the matched filter is replaced
by the decorrelator or the minimum mean square error (MMSE)
receiver as shown in \cite{Meshkati03}. Nash equilibrium is a set
of strategies such that no user can unilaterally improve its own
utility (which is a measure of user satisfaction) given the
strategies taken by other users. Extension of the above results to
multiantenna systems are also presented in \cite{Meshkati03b}.

Multi-carrier CDMA, which combines the benefits of orthogonal
frequency division multiplexing (OFDM) with those of CDMA, is
considered to be a potential candidate for next generation high
data-rate wireless systems (see \cite{HaraPrasad97}). In this
work, we consider an orthogonal multi-carrier DS-CDMA system and
propose a non-cooperative power control game in which each user
tries to choose its transmit power over each carrier to maximize
its overall utility. For the proposed game, we will address the
following questions. When does a Nash equilibrium exist? What kind
of carrier allocations among the competing users will occur at a
Nash equilibrium? Will there be an even spread of usage of the
carriers among users? How does this approach compare with an
approach in which each user simply optimizes over each carrier
independently?

The rest of this paper is organized as follows. In
Section~\ref{background}, we provide some background for this work
by discussing the power control game for the single-carrier case.
The power control game for multi-carrier systems is presented in
Section~\ref{NCPCG}. The Nash equilibrium and its existence for
the proposed game are discussed in Section~\ref{Nash equilibrium}
and a distributed algorithm for reaching the equilibrium (if it
exists) is presented in Section~\ref{algorithm}. The case of
two-carrier systems is studied in more details in
Section~\ref{special case}. Numerical results and conclusions are
given in Sections~\ref{simulation} and \ref{conclusion},
respectively. Our focus throughout this work is on the uplink,
where user terminals transmit to a common concentration point such
as a cellular base station.

\section{Background}\label{background}

Let us first look at the power control game with a single carrier.
To pose the power control problem as a non-cooperative game, we
first need to define a utility function suitable for data
applications. Most data applications are sensitive to error but
they can tolerate delay. It is clear that a higher signal to
interference plus noise ratio (SIR) level at the output of the
receiver will result in a lower bit error rate and hence higher
throughput. However, achieving a high SIR level requires the user
terminal to transmit at a high power which in turn results in low
battery life. In \cite{GoodmanMandayam00}, the utility function of
a user is defined as the ratio of its throughput to its transmit
power, i.e.
\begin{equation}\label{eq1a}
u_k = \frac{T_k}{p_k} .
\end{equation}
Throughput is the net number of information bits that are
transmitted without error per unit time (sometimes referred to as
goodput). It can be expressed as
\begin{equation}\label{eq1}
    T_k = \frac{L}{M} R_k f(\gamma_k)  ,
\end{equation}
where $L$ and $M$ are the number of information bits and the total
number of bits in a packet, respectively. $R_k$ and $\gamma_k$ are
the transmission rate and the SIR for the $k^{th}$ user,
respectively; and $f(\gamma_k)$ is the efficiency function
representing the packet success rate (PSR), i.e. the probability
that a packet is received without an error. Our assumption is that
if a packet has one or more bit errors, it will be retransmitted.
The efficiency function, $f(\gamma_k)$, is assumed to be
increasing and S-shaped (sigmoidal) with $f(\infty)=1$. We also
require that $f(0)=0$ to ensure that $u_k=0$ when $p_k=0$. These
assumptions are valid in many practical systems. It should be
noted that the throughput $T_k$ in (\ref{eq1}) could be replaced
with any increasing concave function as long as we make sure that
$u_k=0$ when $p_k=0$. A more detailed discussion of the efficiency
function  can be found in \cite{Meshkati_TCOMM} and
\cite{GoodmanMandayam01}. Note that for a sigmoidal efficiency
function, the utility function in (\ref{eq1a}) is quasiconcave in
user's transmit power. This is also true if the throughput in
(\ref{eq1}) is replaced with an increasing concave function.

Based on (\ref{eq1a}) and (\ref{eq1}), the utility function for
user $k$ can be written as
\begin{equation}\label{eq2}
    u_k = \frac{L}{M} R_k \frac{f(\gamma_k)}{p_k} .
\end{equation}
This utility function, which has units of \emph{bits/Joule},
captures very well the tradeoff between throughput and battery
life and is particularly suitable for applications where saving
power is as important as achieving a high throughput.

Power control is modeled as a non-cooperative game in which each
user tries to selfishly maximize its own utility. It is shown in
\cite{Saraydar02} that, when matched filters are used as the
uplink receivers, if user terminals are allowed to choose only
their transmit powers for maximizing their utilities, then there
exists an equilibrium point at which no user can improve its
utility given the power levels of other users (Nash equilibrium).
The equilibrium is achieved when users' transmit powers are
SIR-balanced with $\gamma^*$, the solution to $f(\gamma) = \gamma
\ f'(\gamma)$, as the output SIR. Furthermore, this equilibrium is
unique.

In this work, we extend this game-theoretic approach to
multi-carrier systems.

\section{The Non-Cooperative Power Control Game in Multi-carrier Systems} \label{NCPCG}

Let us consider the uplink of a multi-carrier DS-CDMA data network
with $K$ users, $D$ carriers and processing gain $N$ (for each
carrier). The carriers are assumed to be sufficiently far apart so
that the (spread-spectrum) signal transmitted over each carrier
does not interfere with the signals transmitted over other
carriers. At the transmitter, the incoming bits for user $k$ are
divided into $D$ parallel streams and each stream is spread using
the spreading code of user $k$. The $D$ parallel streams are then
sent over the $D$ (orthogonal) carriers. For the $\ell^{th}$
carrier, the received signal at the uplink receiver (after
chip-matched filtering and sampling) can be represented by an
$N\times1$ vector as
\begin{equation}\label{eq4}
    {\mathbf{r}}_{\ell} = \sum_{k=1}^{K} \sqrt{p_{k\ell} h_{k\ell}} \ b_{k\ell}\ {\mathbf{s}}_{k} +
    {\mathbf{w}}_{\ell}
\end{equation}
where $b_{k\ell}$, $p_{k\ell}$, $h_{k\ell}$ are the $k^{th}$
user's transmitted bit, transmit power and path gain,
respectively, for the $\ell^{th}$ frequency channel (carrier).
${\mathbf{s}}_{k}$ is the spreading sequence for user $k$ which is
assumed to be random with unit norm; and ${\mathbf{w}}_{\ell}$ is
the noise vector which is assumed to be Gaussian with mean
$\mathbf{0}$ and covariance $\sigma^2 \mathbf{I}$. The matched
filter output SIR for the $\ell^{th}$ carrier of the $k^{th}$ user
is, therefore, given by
\begin{equation}\label{eq5}
\gamma_{k\ell} = \frac{p_{k\ell} h_{k\ell} } {\sigma^2 +
\frac{1}{N}\sum_{j\neq k} p_{j\ell} h_{j\ell} }\ .
\end{equation}

We propose a non-cooperative game in which each user chooses its
transmit powers over the $D$ carriers to maximize its overall
utility. In other words, each user (selfishly) decides how much
power to transmit over each frequency channel (carrier) to achieve
the highest overall utility. Let $G_D=[{\mathcal{K}}, \{A_k^{MC}
\}, \{u_k^{MC} \}]$ denote the proposed non-cooperative game where
${\mathcal{K}}=\{1, ... , K \}$, and $A_k^{MC}=[0,P_{max}]^D$ is
the strategy set for the $k^{th}$ user. Here, $P_{max}$ is the
maximum transmit power. Each strategy in $A_k^{MC}$ can be written
as ${\mathbf{p}_k} = [p_k^1 , ... , p_k^D]$. The utility function
for user $k$ is defined as the ratio of the total throughput over
the total transmit power for the $D$ carriers, i.e.
\begin{equation}\label{eq6}
    u_k^{MC} = \frac{\sum_{\ell=1}^D T_{k\ell}} {\sum_{\ell=1}^D p_{k\ell}} ,
\end{equation}
where $T_{k\ell}$ is the throughput achieved by user $k$ over the
$\ell^{th}$ carrier, and is given by  $T_{k\ell} = \frac{L}{M} R_k
f(\gamma_{k\ell})$. Hence, the resulting non-cooperative game can
be expressed as the following maximization problem:
\begin{equation}\label{eq7}
    \max_{{\mathbf{p}}_k} \ u_k^{MC} =  \max_{p_{k1} , ... , p_{kD} } u_k^{MC}  \ \ \ \
    \textrm{for}
   \ \  k=1,...,K  ,
\end{equation}
under the constraint of non-negative powers (i.e. $p_{k\ell} \geq
0$ for all $k=1, ... , K$ and $\ell=1, ... , D$). Without any loss
of generality, if we assume equal transmission rates for all
users, (\ref{eq7}) can be expressed as
\begin{equation}\label{eq8}
    \max_{p_{k1} , ... , p_{kD} } \frac{ \sum_{\ell=1}^D f(\gamma_{k\ell})} { \sum_{\ell=1}^D p_{k\ell}} \ \ \ \
    \textrm{for}
   \ \  k=1,...,K  .
\end{equation}
The relationship between $\gamma_{k\ell}$ and $p_{k\ell}$'s is
given by (\ref{eq5}).

\section{Existence of Nash Equilibrium for the Proposed
Game}\label{Nash equilibrium}

For the non-cooperative power control game proposed in the
previous section, a Nash equilibrium is a set of power vectors,
${\mathbf{p}}_1^* , ... , {\mathbf{p}}_K^*$, such that no user can
unilaterally improve its utility by choosing a different power
vector; i.e. ${\mathbf{p}}_1^*, ... , {\mathbf{p}}_K^*$ is a Nash
equilibrium if and only if
\begin{equation}\label{eq9}
   u_k^{MC}({\mathbf{p}}_k^*, {{\mathbf{P}}}_{-k}^*) \geq u_k^{MC}({\mathbf{p}}_k,
    {{\mathbf{P}}}_{-k}^*) \ \textrm{for all} \ {\mathbf{p}}_k ,
\end{equation}
and for $k=1,\cdots,K$; where ${\mathbf{P}}_{-k}^*$ contains the
transmit power vectors of all the users except for user $k$.

Let us define
\begin{equation}\label{eq10}
    \hat{h}_{k\ell} =\frac{ h_{k\ell} } {\sigma^2 +
\frac{1}{N}\sum_{j\neq k} p_{j\ell} h_{j\ell} }
\end{equation}
as the ``effective channel gain" for user $k$ over the $\ell^{th}$
carrier. Based on (\ref{eq5}) and (\ref{eq10}), we have $\gamma_{k
\ell}= \hat{h}_{k \ell} p_{k \ell}$.

\vspace{0.3cm}
\textbf{Proposition 1}: For user $k$, the utility function in
(\ref{eq6}) is maximized when
\begin{equation}\label{eq11a}
p_{k\ell}=\left\{%
\begin{array}{ll}
    p_k^* &  \textrm{for} \ \ \ \ell=L_k\\
    0     &  \textrm{for} \ \ \ \ell \neq L_k \\
\end{array}%
\right. \end{equation} where $L_k=\arg \max_{\ell} \hat{h}_{k
\ell}$ and $p_k^*= \min \{ \frac{\gamma^*}{\hat{h}_k^{L_k}} ,
P_{max} \}$, with $\gamma^*$ being the unique (positive) solution
of $f(\gamma) = \gamma \ f'(\gamma)$.\vspace{0.2cm}

\emph{Proof}: We first show that $\frac{f(\tilde{a} p)}{p}$ is
maximized when ${p=\gamma^*/\tilde{a}}$. For this, we take the
derivative of $\frac{f(\tilde{a} p)}{p}$ with respect to $p$ and
equate it to zero. Letting $\gamma= \tilde{a} p$, we have
\begin{equation}\label{eq11}
    p  \frac{\partial \gamma}{\partial p}  f'(\gamma) - f(\gamma)
    =0 .
\end{equation}
Since $p  \frac{\partial \gamma}{\partial p} =\gamma$, $p^*=\min
\{\gamma^*/\tilde{a}, P_{max} \}$ maximizes $\frac{f(\tilde{a}
p)}{p}$, where $\gamma^*$ is the unique (positive) solution to
$f(\gamma)=\gamma f'(\gamma)$. Now, without any loss of
generality, let us assume that ${\hat{h}_{k1} > \hat{h}_{k2} > ...
> \hat{h}_{kD}}$. Then, based on the above argument we can write
$\frac{f(\hat{h}_{k1} p_{k1})}{p_{k1}} \leq \frac{f(\hat{h}_{k1}
p_k^*)}{p_k^*}$ for any $p_{k1} \geq 0$, where $p_k^*= \gamma^*/
\hat{h}_{k1}$. Also, because $\hat{h}_{k1} > \hat{h}_{k2} > ... >
\hat{h}_{kD}$, we have $\frac{f(\hat{h}_{k\ell}
p_{k\ell})}{p_{k\ell}} \leq \frac{f(\hat{h}_{k1} p_k^*)}{p_k^*}$
 for all $p_{k\ell} \geq 0$ and $\ell=2,3, ..., K$. Based on the
above inequalities, we can write
\begin{equation}\label{eq12}
\frac{f(\hat{h}_{k \ell} p_{k \ell})} {f(\hat{h}_k^1 p_k^*)}
 \leq \frac{p_{k \ell}}{p_k^*} , \ \ \ \ \  \textrm{for} \ \ \ell=1, 2,
 ... , D .
\end{equation}

Adding the $D$ inequalities given in (\ref{eq12}) and rewriting
the resulting inequality, we have
\begin{equation}\label{eq13}
 \frac{ \sum_{\ell=1}^D f(\hat{h}_{k\ell} p_{k\ell})} { \sum_{\ell=1}^D p_{k\ell}}
\leq \frac{ f( \hat{h}_{k1} p_k^*)} {p_k^*}  \ \ \textrm{for all}
\ \ p_{k1}, ... , p_{kD} \geq 0 \ .
\end{equation}
This completes the proof. \ ${\Box}$ \vspace{0.3cm}

Proposition 1 suggests that the utility for user $k$ is maximized
when the user transmits only over its ``best" carrier such that
the achieved SIR at the output of the matched filter receiver is
equal to $\gamma^*$. The ``best" carrier is the one with the
largest effective channel gain. An alternative way of interpreting
Proposition 1 is that the utility for user $k$ is maximized when
the user transmits only on the carrier that requires the least
amount of transmit power to achieve $\gamma^*$ at the output of
the uplink receiver. A set of power vectors, ${\mathbf{p}}_1^* ,
... , {\mathbf{p}}_K^*$, is a Nash equilibrium if and only if they
simultaneously satisfy (\ref{eq11a}).

It should also be noted that the utility-maximizing strategy
suggested by Proposition 1 is different from the waterfilling
approach discussed in \cite{WeiYu02}. This is because in
\cite{WeiYu02}, utility is defined as the user's throughput
whereas here we have defined utility as the number of bits
transmitted per joule of energy consumed. Another observation is
that if we use $\tilde{u}_k= \sum_{\ell=1}^D
\frac{T_{k\ell}}{p_{k\ell}}$  as the utility function, then the
utility for user $k$ is maximized when the user transmits on all
the carriers at power levels that achieve $\gamma^*$ for every
carrier. This is equivalent to the case in which each user
maximizes its utility over each carrier independently.

Since at Nash equilibrium (if it exists), each user must transmit
on one carrier only, there are exactly $D^K$ possibilities for an
equilibrium. For example, in the case of $K=D=2$, there are 4
possibilities for Nash equilibrium:
\begin{itemize}
    \item User 1 and user 2 both transmit on the first carrier.
    \item User 1 and user 2 both transmit on the second carrier.
    \item User 1 transmits on the first carrier and user 2 transmits on the second carrier.
    \item User 1 transmits on the second carrier and user 2 transmits on the first carrier.
\end{itemize}

Depending on the set of channel gains, i.e. $h_{k\ell}$'s , the
proposed power control game may have no equilibrium, a unique
equilibrium, or more than one equilibrium. Let us for now assume
that the processing gain is sufficiently large so that even when
all $K$ users transmit on the same carrier, $\gamma^*$ can be
achieved by all users. This is the case when $N>(K-1)\gamma^*$.
The following proposition helps identify the Nash equilibrium (if
it exists) for a given set of channel gains. \vspace{0.3cm}

\textbf{Proposition 2}: The necessary condition for user $k$ to
transmit on the $\ell^{th}$ carrier at equilibrium is that
\begin{equation}\label{eq14}
    \frac{ h_{k \ell}} {h_{ki}} > \frac{\Theta_{n(\ell)} } {
    \Theta_{n(i)} } \ \Theta_0  \ \ \ \ \  \textrm{for all}\ i\neq
    \ell \ ,
\end{equation}
where $n(i)$ is the number of users transmitting on the $i^{th}$
carrier and
\begin{equation}\label{eq15}
    \Theta_n = \frac{1} { 1 - (n-1) \frac{\gamma^*}{N}} \ \ \ \ n=0,
    1, ... , K .
\end{equation}
In this case, $p_{k\ell}= \frac{\gamma^* \sigma^2}{h_{k\ell}}
 \Theta_{n(\ell)}$.\vspace{0.2cm}

\emph{Proof}: Based on Proposition 1, in order for user $k$ to
transmit on carrier $\ell$ at equilibrium, we must have
\begin{equation}\label{eq15a}
\hat{h}_{k \ell}
> \hat{h}_k^{i} \ \textrm{for all} \ i \neq \ell .
\end{equation}
Since $n(\ell)$ users (including user $k$) are transmitting on the
$\ell^{th}$ carrier and $n(i)$ users are transmitting on the
$i^{th}$ carrier and all users have an output SIR equal to
$\gamma^*$, we have
\begin{equation}\label{eq15b}
\hat{h}_{k\ell} =\frac{ h_{k\ell} } {\sigma^2 + \frac{
n(\ell)-1}{N} q_{\ell} } ,
\end{equation}
and
\begin{equation}\label{eq15c}
\hat{h}_{ki} =\frac{ h_{ki} } {\sigma^2 + \frac{n(i)}{N} q_i } ,
\end{equation}
where $q_{\ell}=\frac{\sigma^2 \gamma^*}{1- \frac{( n(\ell)-1)
\gamma^*}{N}}$ and
 $q_i=\frac{\sigma^2 \gamma^*}{1- \frac{( n(i)-1)
\gamma^*}{N}}$ are the received powers for each user on the
$\ell^{th}$ and $i^{th}$ carriers, respectively. Now define
$\Theta_n= \frac{1}{1-(n-1)\frac{\gamma^*}{N}}$ to get $q_{\ell}=
\sigma^2 \gamma^* \Theta_{n(\ell)}$ and $q_i=\sigma^2 \gamma^*
\Theta_{n(i)}$. Substituting $q_{\ell}$ and $q_i$ into
(\ref{eq15b}) and (\ref{eq15c}) and taking advantage of the fact
that $1+\frac{(n-1)\gamma^*}{N}\Theta_n = \Theta_n$ we get
\begin{equation}\label{eq15d}
\hat{h}_{k\ell} =\frac{ h_{k\ell} } {\Theta_{n(\ell)} } ,
\end{equation}
and
\begin{equation}\label{eq15e}
\hat{h}_{ki} =\frac{ h_{ki} } { \frac{\Theta_{n(i)}}{\Theta_0} } .
\end{equation}
Consequently, (\ref{eq14}) is obtained by substituting
(\ref{eq15d}) and (\ref{eq15e}) into (\ref{eq15a}). Furthermore,
since $p_{k\ell} h_{k\ell} = q_{\ell} =\sigma^2 \gamma^*
\Theta_{n(\ell)}$, we have  $p_{k\ell}= \frac{\gamma^*
\sigma^2}{h_{k\ell}} \Theta_{n(\ell)}$ and this completes the
proof. \ ${\Box}$ \vspace{0.3cm}

Based on (\ref{eq15}), when $N > (K-1) \gamma^*$, we have ${0<
\Theta_0 < \Theta_1 < \Theta_2 < ... < \Theta_K}$ with
$\Theta_1=1$.

For each of the $D^K$ possible equilibria, the channel gains for
each of the $K$ users must satisfy $D-1$ inequalities similar to
(\ref{eq14}). Furthermore, satisfying a set of $K(D-1)$ of such
inequalities by the $K$ users is sufficient for existence of Nash
equilibrium but the uniqueness is not guaranteed. For example, for
the case of $K=D=2$, the 4 possible equilibria can be
characterized as follows.
\begin{itemize}
    \item For both users to transmit on the first carrier at equilibrium, we must have
      $\frac{h_{11}}{h_{12}}>\Theta_2$ and $\frac{h_{21}}{h_{22}}>\Theta_2$.
    \item For both users to transmit on the second carrier at equilibrium, we must have
      $\frac{h_{12}}{h_{11}}>\Theta_2$ and $\frac{h_{22}}{h_{21}}>\Theta_2$.
    \item For user 1 and user 2 to transmit on the first and second carriers, respectively, at equilibrium, we must have
     ${\frac{h_{11}}{h_{12}}>\Theta_0}$ and $\frac{h_{22}}{h_{21}}>\Theta_0$.
    \item For user 1 and user 2 to transmit on the second and first carriers, respectively, at equilibrium, we must have
      ${\frac{h_{12}}{h_{11}}>\Theta_0}$ and $\frac{h_{21}}{h_{22}}>\Theta_0$.
\end{itemize}

\begin{figure}
\centering
\includegraphics[width=8.5cm]{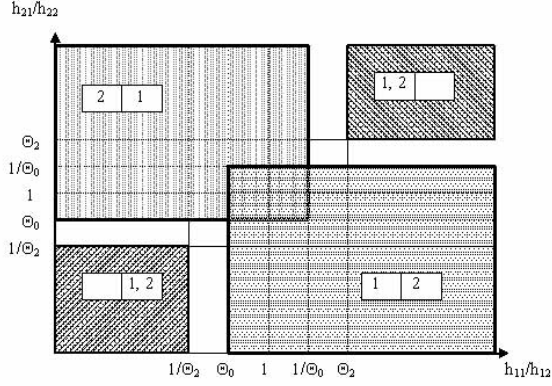}
\caption{Nash Equilibrium Regions for the Case of $K=2$ and $D=2$}
\label{fig2}
\end{figure}

Fig. \ref{fig2} shows the regions corresponding to the above four
equilibria. It can be seen that for certain values of channel
gains there is no Nash equilibrium and for some values of channel
gains there are two equilibria.

\section{A Distributed Algorithm}\label{algorithm}
In this section, we present an iterative and distributed algorithm
for reaching the Nash equilibrium (if it exists) for the proposed
power control game. The description of the algorithm is as
follows.
\begin{enumerate}
    \item Let $k=1$
    \item Given the transmit powers of other users, user $k$ picks
    the carrier whose effective channel gain (i.e. $\hat{h}_{k \ell}$)
    is the largest and transmits only on that carrier at a power
    level equal to $p_k^*$
    \item $k=k+1$
    \item If $k \leq K$ then go back to step 2
    \item Stop if the powers have converged, otherwise go to {step~1}
\end{enumerate}

It is clear that if the above algorithm converges, it will
converge to a Nash equilibrium. It is also demonstrated in Section
\ref{simulation} that the algorithm converges when a Nash
equilibrium exists. In the case of multiple Nash equilibria, the
above algorithm converges to one of the equilibria depending on
the starting point.

\section{Special Case of Two Carriers}\label{special case}

To gain some insight into the properties of the Nash equilibria
for our proposed game, let us concentrate on a system with two
carriers and two users (i.e. $D=K=2$). We assume that the channel
amplitudes are independent and identically distributed (i.i.d.)
among users and carriers. If we assume Rayleigh fading (with mean
1) for the channel amplitudes then the channel gains,
$h_{k\ell}$'s, will be i.i.d. with exponential distribution of
mean 1. Let $X_1$ be the random variable corresponding to the
number of users that transmit over the first carrier at
equilibrium. If we assume $N>\gamma^*$, then, based on Proposition
2, the probability that both users transmit on the first carrier
at equilibrium (i.e. $X_1=2$) is given by
\begin{eqnarray}\label{eq16}
    P_{X_1}(2)&=&Pr\{X_1=2\} \nonumber\\
    &=& Pr\left\{\frac{h_{11}}{h_{12}}>\Theta_2, \frac{h_{21}}{h_{22}}>\Theta_2\right\}
    =\left(\frac{1}{1+\Theta_2}\right)^2.\hspace{0.5cm}
\end{eqnarray}
Similarly, the probability of both users transmitting on the
second carrier at equilibrium is
\begin{eqnarray}\label{eq16b}
    P_{X_1}(0)&=&Pr\{X_1=0\}\nonumber\\
    &=& Pr\left\{\frac{h_{12}}{h_{11}}>\Theta_2, \frac{h_{22}}{h_{21}}>\Theta_2\right\}=
    \left(\frac{1}{1+\Theta_2}\right)^2.\hspace{0.5cm}
\end{eqnarray}
The probability of one user transmitting on each of the two
carriers can be found to be
\begin{equation}\label{eq16c}
    P_{X_1}(1)=Pr\{X_1=1\} = 2\left(\frac{1}{1+\Theta_0}\right)^2-\left(\frac{1-\Theta_0}{1+\Theta_0}\right)^2 .
\end{equation}
Consequently, the probability that no Nash equilibrium exists is
given by
\begin{eqnarray}\label{eq16d}
    P_o&=&Pr\{No\ Nash\ equilibrium\} \nonumber\\&=& 2\left\{\left(\frac{\Theta_0}{1+\Theta_0}\right)^2 -
    \left(\frac{1}{1+\Theta_2}\right)^2\right\} .
\end{eqnarray}
It should be noted that as the processing gain $N$ becomes larger,
$\Theta_0$ and $\Theta_1$ approach 1 from below and above,
respectively. This results in a reduction in $P_{X_1}(1)$ but an
increase in $P_{X_1}(0)$ and $P_{X_1}(2)$, i.e. the probability
mass function (pmf) for $X_1$ becomes flatter. However, the
increase outweighs the decrease and as a result $P_o$ decreases as
$N$ increases. Going back to Fig. \ref{fig2}, we see that the
region for which no Nash equilibrium exists shrinks as $N$
increases. In addition, the region for which more than one
equilibrium exists disappears as $N$ becomes very large.
Therefore, we can say that as the processing gain becomes large,
the probability that the proposed power control game has a unique
Nash equilibrium approaches one.

So far, the assumption has been that $N>\gamma^*$ so that both
users can achieve $\gamma^*$ even when they are transmitting over
the same carrier. For the case of $N \leq \gamma^*$, the users
cannot achieve $\gamma^*$ simultaneously when they are
transmitting on the same carrier and hence they would end up
transmitting at the maximum power. Therefore, when
$N\leq\gamma^*$, the probability of both users transmitting on the
same carrier at equilibrium is virtually zero, i.e. $P_{X_1}(2)=
P_{X_1}(0)\approx 0$. Hence, we have
$P_o=2\left(\frac{\Theta_0}{1+\Theta_0}\right)^2$ .

Although obtaining explicit expressions for the probabilities of
the occurrence of various Nash equilibria for the case of $K>2$ is
difficult, many of the statements made for the case of $K=2$ are
also valid when $K>2$. Namely, as $N$ increases the pmf of $X_1$
becomes wider and at the same time the probability that no
equilibrium exists becomes smaller. This means that in the
asymptotic case of large processing gain, the proposed power
control game has a unique equilibrium. Furthermore, for very large
values of $N$, the pmf of $X_1$ can be approximated as
\begin{equation}\label{eq18}
P_{X_1}(m) = Pr\{X_1=m\}\approx\left(%
\begin{array}{c}
  K \\
  m \\
\end{array}%
\right)\left(\frac{1}{2}\right)^K \textrm{for} \ m=0,\cdots, K.
\end{equation}

In the next section, we show the validity of these claims using
simulation.

\section{Simulation Results} \label{simulation}

We first consider the case of two carriers with two users. We
assume $L=M=100$, $R=100$ Kbps and $\sigma^2=5\times 10^{-16}$
Watts; and use $f(\gamma)=(1-e^{-\gamma})^M$ as the efficiency
function. For this efficiency function, $\gamma^*=6.4$ (=8.1dB).
We assume that the channel gains are i.i.d. with exponential
distribution of mean 1. We consider 20,000 realizations of channel
gains. For each realization, we run the algorithm proposed in
{Section~\ref{algorithm}} for 20 iterations. If convergence is
reached by the end of the $20^{th}$ iteration, we record the
number of users that transmit on each carrier; otherwise, we
assume there is no equilibrium. For our simulations, $P_{max}$ is
assumed to be very large which translates to having no transmit
power limit for the user terminal.

\begin{figure}
\centering
\includegraphics[width=7.25cm]{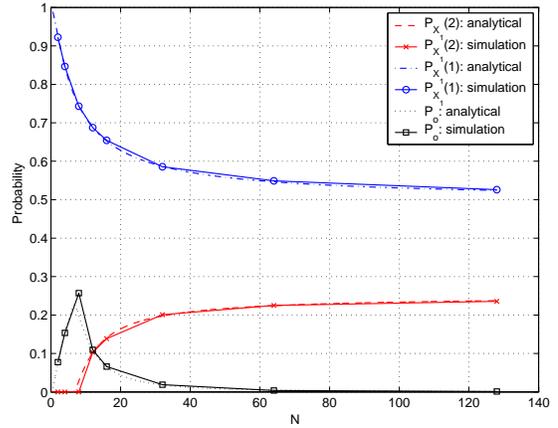}
\caption{Probability of having $m$ users on the first carrier at
equilibrium, $P_{X_1}(m)$, for $m=1,2$ as well as probability of
having no equilibrium are shown for $D=K=2$.}\label{probMF}
\end{figure}

It is observed that the distributed algorithm proposed in Section
\ref{algorithm} converges when a Nash equilibrium exists. Recall
that $P_{X_1}(m)$ represents the probability that $m$ users
transmit on the first carrier at equilibrium. Fig. \ref{probMF}
shows $P_{X_1}(2)$, $P_{X_1}(1)$ and $P_o$ (probability of no
equilibrium) as a function of the processing gain $N$. The
analytical expressions obtained in Section \ref{special case} are
also plotted. We see that there is a close agreement between the
simulation results and the analytical values. It is also observed
that as $N$ becomes large, $P_o$ approaches zero. For $N=16$, for
example, the probability that a Nash equilibrium exists is about
93\%. Since $P_{X_1}(0)$ is identical to $P_{X_1}(2)$, it is not
shown in the figure.

\begin{figure}
\centering
\includegraphics[width=7.25cm]{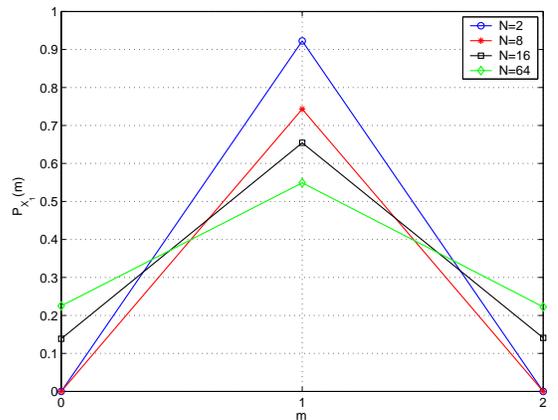}
\caption{Probability mass function of $X_1$ for different
processing gains ($D=2$ and $K=2$)}\label{pmfMF}
\end{figure}

Fig. \ref{pmfMF} shows $P_{X_1}(m)$ as a function of $m$ for
different values of $N$. We can see from the figure that as the
processing gain increases, the pmf of $X_1$ becomes flatter. This
is because for larger values of $N$, the system becomes more
tolerant towards interference. Therefore, the probability with
which the two users are able to transmit on the same carrier at
equilibrium increases.

\begin{figure}[t]
\centering
\includegraphics[width=7.05cm]{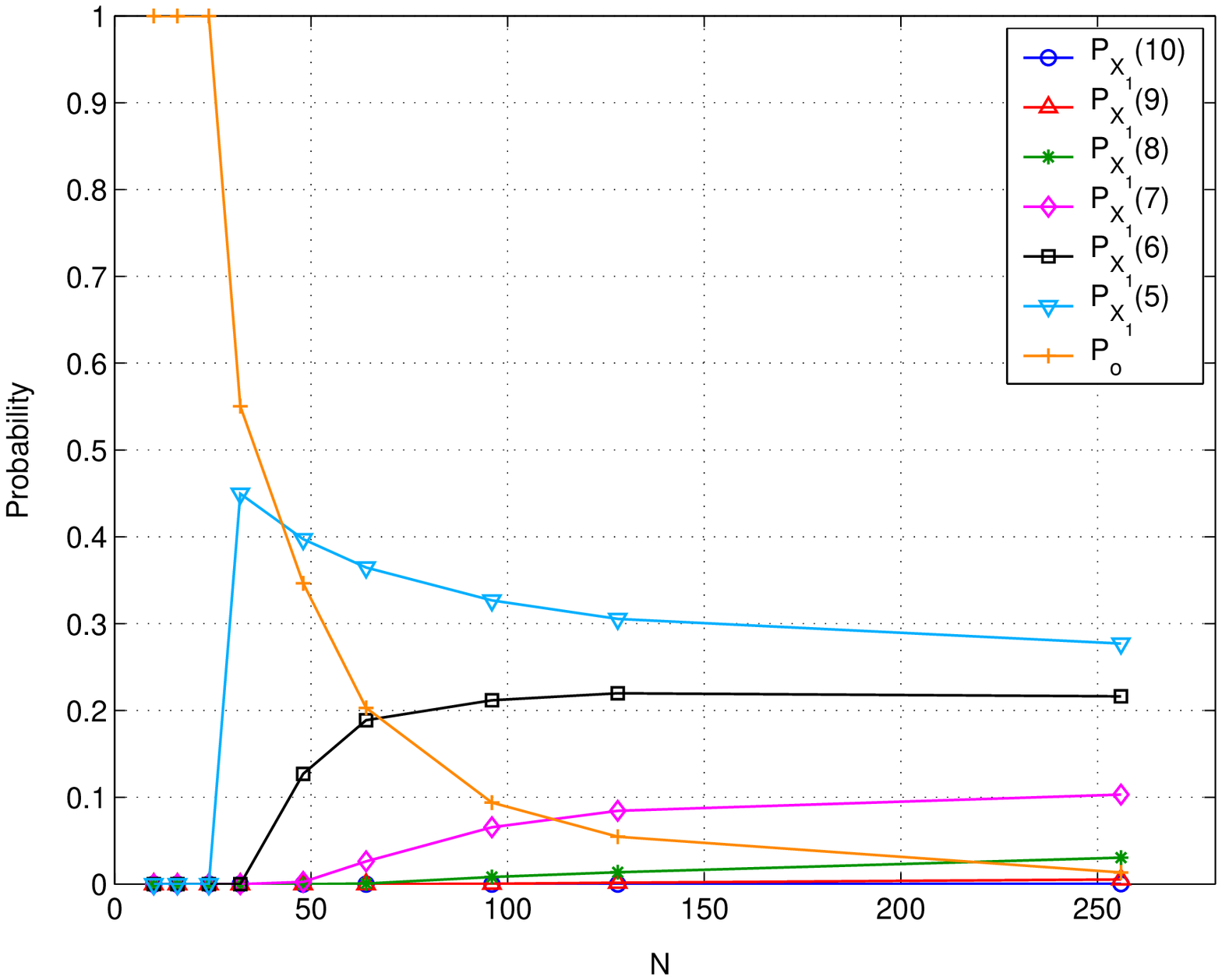}
\caption{Probability of having $m$ users on the first carrier at
equilibrium, $P_{X_1}(m)$, for $m=5, 6,\cdots, 10$ as well as
probability of having no equilibrium are shown for $D=2$ and
$K=10$.}\label{probMF-K10}
\end{figure}

\begin{figure}[t]
\centering
\includegraphics[width=7.25cm]{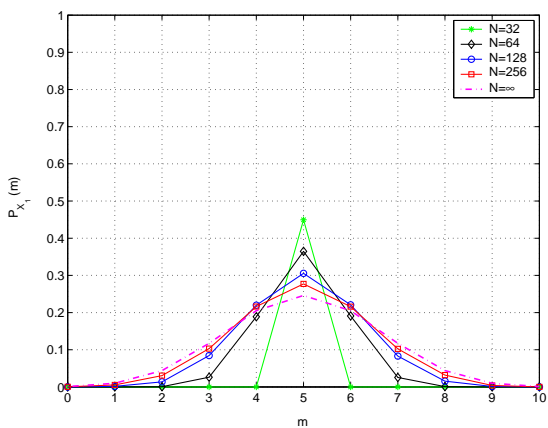}
\caption{Probability mass function of $X_1$ for different
processing gains ($D=2$ and $K=10$)}\label{pmfMF-K10}
\end{figure}

We repeat the above experiment for the case of two carriers and 10
users. Fig. \ref{probMF-K10} shows $P_{X_1}(m)$ as a function of
the processing gain. Due to symmetry, we have only plotted the
probabilities for $m=5,6,...,10$. The probability that no
equilibrium exists ($P_o$) is also shown. Similar trends as those
observed in the case of $K=2$ are also seen here. We observe that
here again as $N$ becomes large, $P_o$ approaches zero.
$P_{X_1}(m)$ as a function of $m$ for different values of $N$ is
plotted in Fig. \ref{pmfMF-K10}. The asymptotic approximation for
$P_{X_1}(m)$ which is given by (\ref{eq18}) is also shown. We see
from the figure that as $N$ increases the pmf of $X_1$ becomes
wider because the system becomes more interference-tolerant. Also,
the equilibria for which the allocation of users to the carriers
is highly asymmetric (e.g. $m=9$ and 10) are unlikely to happen.
In other words, with a high probability, the users are evenly
distributed between the two carriers. It should be noted that when
$N$ is small, $\gamma^*$ cannot be achieved simultaneously by all
the users at the output of the matched filters. Therefore, users
keep increasing their transmit powers and hence no equilibrium is
reached. This is the case until $N$ becomes large enough so that
it can accommodate at least 5 users on each carrier (i.e.
$N>25.6$). In practice, however, if $N$ is not large enough, some
or all users end up transmitting at the maximum power.

\begin{figure}
\centering
\includegraphics[width=7.025cm]{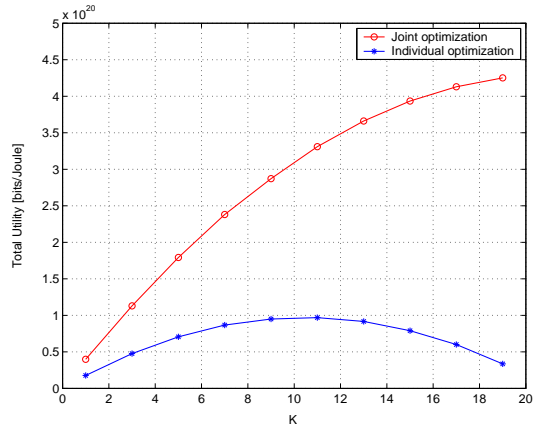}
\caption{Total Utility vs Number of Users ($D=2$ and
$N=128$)}\label{util-compare}
\end{figure}

We now compare the proposed approach, which jointly maximizes each
user's utility over all carriers, with the approach that maximizes
user's utility independently over each carrier. In the joint
maximization approach each user transmits only on the carrier that
has the best effective channel whereas in the other case, all
users transmit on all the carriers such that the output SIR on
each carrier is $\gamma^*$. We consider a system with 2 carriers
and $N=128$. We fix $K$ and compute the sum of the utilities
achieved by all users for 20,000 channel realizations. The utility
for each case is the ratio of the total transmitted bits over the
two carriers divided by the total energy consumed. Fig.
\ref{util-compare} shows the average total utility vs. $K$ for the
two approaches. We see a significant improvement in the achieved
utility when joint maximization over all carriers is used. This is
because when all the users transmit on every carrier, they cause
unnecessary interference. To achieve $\gamma^*$, each user is
hence forced to transmit at a higher power level which in turn
results in a considerable reduction in the overall utility.

\section{Conclusion} \label{conclusion}
We have modelled power control for multi-carrier CDMA systems as a
non-cooperative game in which each user needs to decide how much
power to transmit over each carrier to maximize its overall
utility. The utility function has been defined as the overall
throughput divided by the total transmit power over all the
carriers and has units of bits per joule. For this utility
function, we have shown that the utility is maximized when each
user transmits only on the carrier which has the best ``effective
channel" for that user. In addition, we have derived the
conditions for existence of a Nash equilibrium and proposed an
iterative and distributed algorithm for reaching the equilibrium
(if it exists). The properties of the Nash equilibria for the
proposed game have been studied.  We have also demonstrated the
performance improvement resulting from the proposed approach
compared to the approach of independent optimization over each
carrier.


\end{document}